\documentclass[journal,onecolumn]{IEEEtran}
\usepackage{caption}
\usepackage{subcaption}
\usepackage{graphicx}

\ifCLASSINFOpdf
\else
\fi
\begin{document}
%
\title{Blockchain Meets AI for  Resilient and Intelligent Internet of Vehicles}
%
%
%

\author{Pranav~K.~Singh, Sukumar~Nandi,	Sunit~K.~Nandi, Uttam Ghosh and Danda~B.~Rawat
\thanks{The work is published in IEEE COMSOC MMTC Communications - Frontiers Vol. 16, No. 6, November 2021, Page 12-24}}

\maketitle

\begin{abstract}
The Internet of Vehicles (IoV) is flourishing and offers various applications relating to road safety, traffic and fuel efficiency, and infotainment. Dealing with security and privacy threats and managing the trust (detecting malicious and misbehaving peers) in IoV remains the most significant concern. Artificial Intelligence is one of the most revolutionizing technologies, and the predictive power of its machine learning models can help detect intrusions and misbehaviors. Similarly, empowering the state-of-the-art IoV security framework with blockchain can make it secure and resilient. This article discusses joint AI and blockchain for security, privacy and trust-related risks in IoV. This paper also presents problems, challenges, requirements and solutions using ML and blockchain to address aforementioned issues in IoV.
\end{abstract}


%
\IEEEpeerreviewmaketitle

\section{INTERNET OF VEHICLES (IOV) SYSTEMS OVERVIEW}
%
%
%
%

\subsection{Overview}
\begin{figure}[!h]
	\centering
	\includegraphics[width=0.7\textwidth]{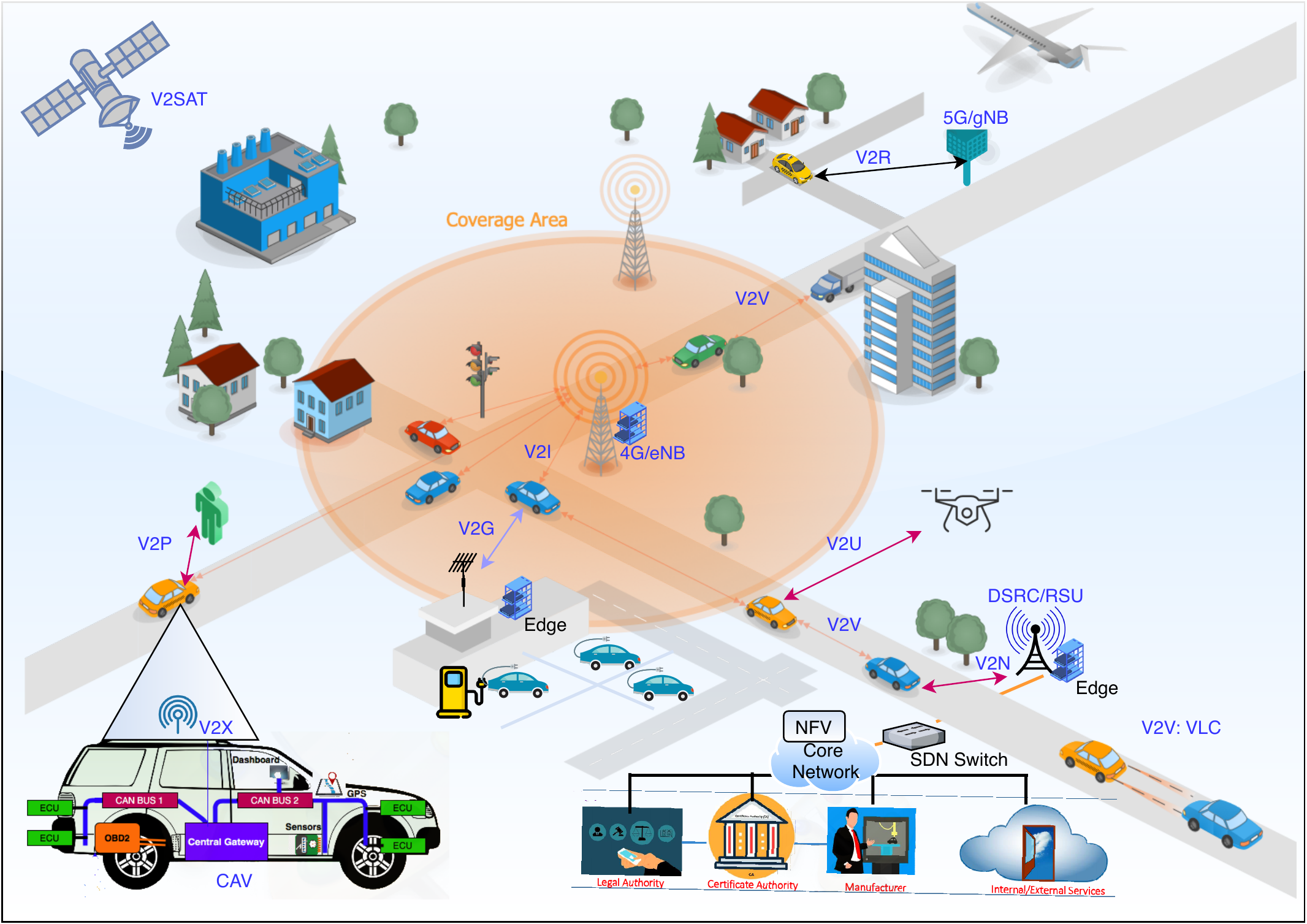}
	\caption{An Overview of Internet-of-Vehicles}
	\label{fig:0}
\end{figure}

In the new era of the Internet of Everything (IoE), the traditional VANETs enabled by vehicle-to-vehicle (V2V) and vehicle-to-infrastructure (V2I) communication have evolved to the Internet of Vehicles (IoV). IoV is the concept that connects smart and intelligent vehicles to any other entities around them, such as vehicles, infrastructure, pedestrians, networks, grids, UAVs, etc., through vehicle-to-everything (V2X) or cellular V2X (C-V2X) technologies \cite{abou2019cellular}. The generic view of IoV is shown in Fig. \ref{fig:0}. 

\subsection{Applications}
IoV is expected to solve the major challenges of our transportation by improving road safety, minimizing road congestion, reducing fuel consumption and CO2 emissions, solving parking issues, and minimizing expenses and space by enabling cab sharing, etc. The remarkable advancements in on-board capabilities of vehicles (sensing, computation, storage, communication), radio access technologies (RAT), network architectures, protocol stacks have enabled automated driving and platooning. These advancements brought IoV to the center of Industry 4.0. and led to promising areas of intelligent transportation, vehicle manufacturing, payment services, predictive maintenance, usage-based insurance, intelligent parking, automation, infotainment, software, energy, secure data sharing, data trading, vehicle life-cycle, etc. \cite{contreras2017internet}. 

\subsection{Key Enablers}
To support a wide variety of IoV applications and services, some of the best-known RATs are DSRC / IEEE 802.11p, Wi-Fi, 4G, new 5G (NR) radio, 6G (envisioned) \cite {tang2019future}, white space TV, millimeter wave (mmWave) and visible light communications (VLC). Three well-known regional protocol stacks developed over DSRC for vehicular communications are Connected Vehicle (WAVE), the Cooperative-ITS (C-ITS), and the ARIB STD-T109 in the USA, Europe, and Japan, respectively \cite{singh2019tutorial}. However, it is soon realized that a single RAT (DSRC) and traditional solutions at core may not support diverse QoS requirements of emerging applications. For instance, these regional protocol stacks over DSRC may not be sufficient to provide the end-to-end latency and throughput requirements of Autonomous Vehicles (AVs). Thus, various technologies are being developed and tested to handle the highly dynamic (variable density, high mobility, diversified QoS) and complex IoV. The ultimate goal is to provide secure, seamless, ultrareliable low latency connectivity and support high throughput and high capacity. 

Artificial Intelligence (AI) s a promising approach for making IoV intelligent. AI is implemented using machine learning (supervised, unsupervised, reinforcement) models. AI integration into IoV can help improve resource allocation and management, decision making, and detect misbehaviors, anomalies, and intrusions. Blockchain is the next revolutionizing technology \cite{puthal2018blockchain}, which can be integrated as a decentralized security framework in the IoV. Leveraging Blockchain into IoV can improve security, privacy, and trust. It can secure the cyber-physical system (CPS) of connected and autonomous vehicles (CAVs), boost PKI capabilities, and enhance the scalability and availability of the IoV security framework. The inherent characteristics of Blockchain can also be capitalized to enable a wide variety of services in IoV, such as secure data sharing and trading among two or more entities.

\section{AI and Blockchain for IoV}

In this section, we discuss major IoV challenges that AI and blockchain can solve. 

\subsection{AI for IoV}

\begin{figure*}
	\begin{subfigure}[b]{0.48\textwidth}
		\includegraphics[width=\textwidth]{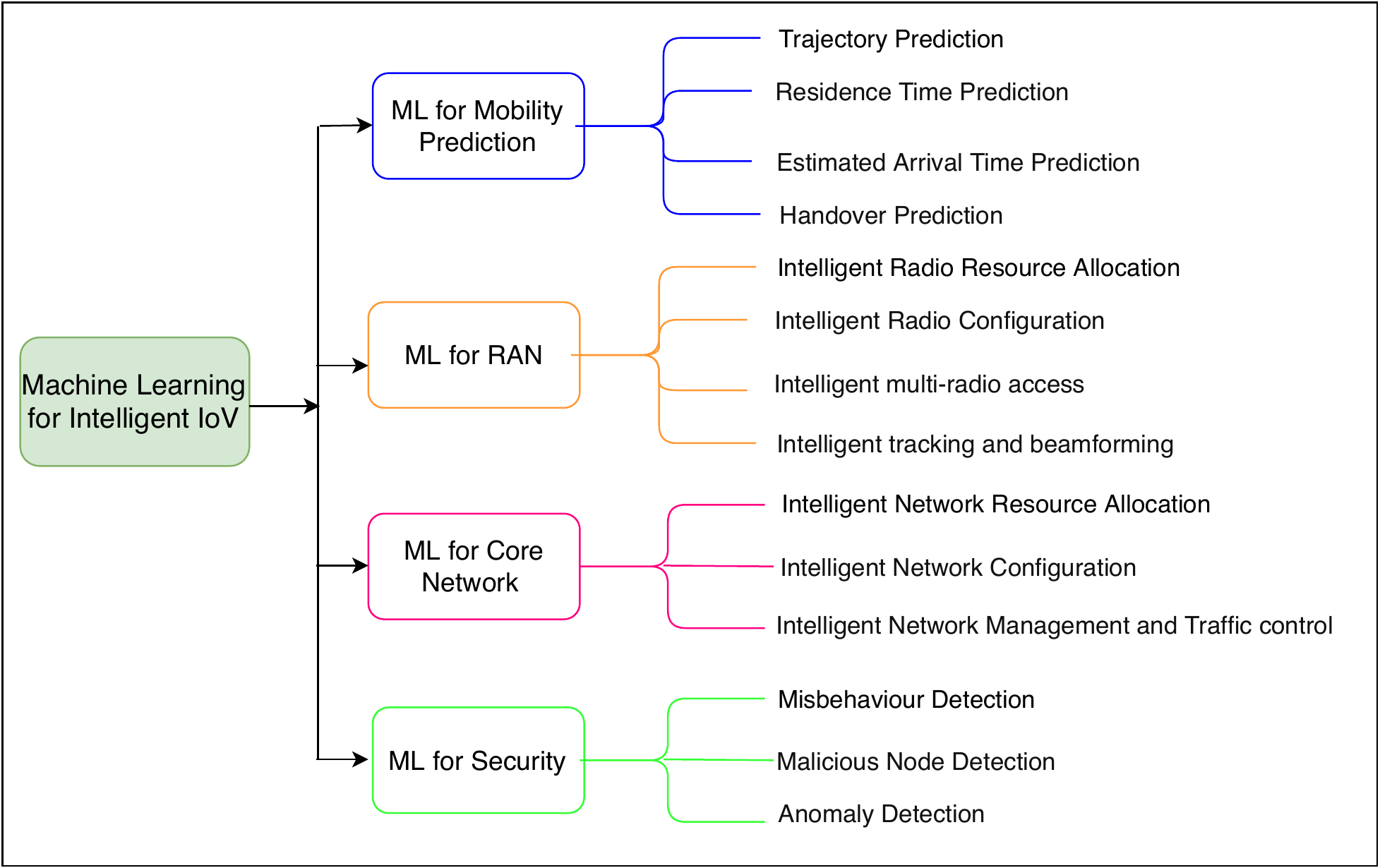}
		\caption{AI/Machine Learning for Intelligent IoV}
		\label{fig:2}
	\end{subfigure}
	\hfill
	\begin{subfigure}[b]{0.50\textwidth}
		\centering
		\includegraphics[width=\textwidth]{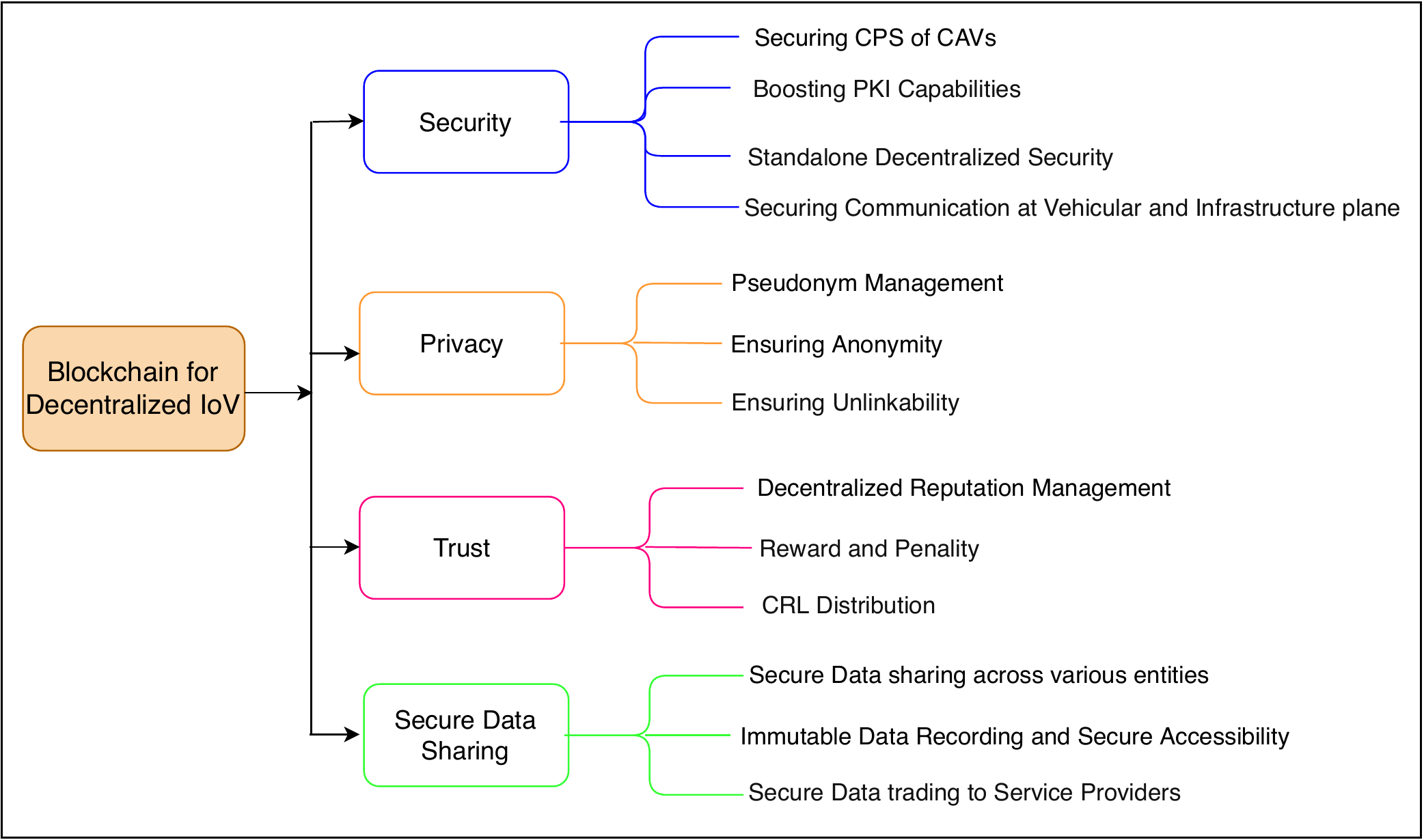}
		\caption{Blockchain for Decentralized IoV}
		\label{fig:3}
	\end{subfigure}
	\caption{AI and Blockchain for IoV}
\end{figure*}

Fig. \ref{fig:2} lists major IoV challenges such as mobility management, radio, and network resource management, and security \cite{tang2019future}, where AI/ML can play a vital role. 

Finding the vehicle's future position in IoV is challenging because of its highly dynamic nature. Mobility prediction can play a crucial role in various applications and services for IoV. AI and ML can help in an accurate mobility prediction, facilitating better handover and radio resource management and avoiding degradation of the expected QoE and QoS. 

The conventional techniques in the radio access network (RAN) may not satisfy the high throughput and ultralow-delay requirement of IoV. Thus, AI/ML techniques are being explored to address the challenges associated with conventional radio resource allocation, radio configuration, multi-radio access, tracking, and beamforming.

The HetNet scenario results in new challenges at the core network of IoV in terms of network resource allocation, configuration, management, and traffic control. The AI/ML techniques are being widely used in IoV for network radio resource allocation, transmission power control, beamforming, load balancing, scheduling, traffic offloading, admission control, dynamic routing, etc.

The cyber-physical systems of smart and connected vehicles and V2X communication are susceptible to various hacks and attacks. Major attacks are denial-of-service (DoS), Replay, Sybil, man-in-middle, bogus and fake information propagation, wormhole, blackhole attacks, etc. The malicious behaviors of IoV users are another big challenge. Although various solutions (PKI, group signature) have been proposed, the detection of malicious behavior, anomaly, and intrusions, and prevention from the mentioned attacks remains the biggest challenge to be addressed. Various deep learning models are being explored to detect misbehavior and attacks and help the trusted authority take preventive measures.

\subsection{Blockchain for IoV}

IoV ecosystem (in-vehicle (sensors), communication (V2X), control, applications, and services) faces several challenges in terms of security, privacy, trust management, data, and resource sharing and trading. Fig. \ref{fig:3} lists IoV challenges related to security, privacy, trust, and secure data sharing where blockchain can help. 

\subsubsection{Blockchain to solve the challenges of IoV} The list of IoV challenges that blockchain can solve are as follows:

\paragraph{Security, Privacy, and Trust} Integration of sensors and vehicle-to-everything connectivity expands attack vectors for malicious entities. Extensive data sharing and open wireless broadcast attract adversaries to exploit privacy. IoV users with malicious intents exploit existing security flaws to gain over others. The blockchain integration into IoV with smart contracts and an advanced cryptographic approach can address these challenges, improve security, preserve privacy, and build trust.

\paragraph{Availability and Fault Tolerance} 
In a state-of-the-art centralized IoV system, tackling the single point of failure and ensuring data and service availability are among the most significant challenges. The distributed and decentralized nature of blockchain eliminates the reliance on central servers or clouds and allows replication of records among other peers. Thus, blockchain adoption can make IoV fault-tolerant, and resilient \cite{tripathi2019role}. 

\paragraph{Data Sharing} The CAVs generate massive data, which must be shared securely among peers and service providers for safety, traffic, and other services. However, the state-of-the-art data sharing mechanism lacks security, user privacy, reliability, trust, and efficiency.  Blockchain has proven its strength to build trust and reliability in similar topologies and is capable of resolving other limitations of secure and private data sharing of IoV.

\paragraph{Data and Resource Trading} IoV provides excellent business opportunities for vehicle owners and service providers by enabling the trade of data and resources. However, the existing system does not facilitate such a platform for data and resource trade, which is fair, secure, transparent, and preserves the user's privacy. Blockchain has the potentials to provide a secure, trusted, fair, and decentralized platform for data and resource trading among various IoV entities. 

\paragraph{Traffic Monitoring and Control} 
The trafﬁc monitoring (traffic condition, rule violation),  automated trafﬁc control \& management rely on data generated by vehicles uploaded to RSUs and edges for storage. However, they are vulnerable to attacks related to data availability, data integrity, and user privacy. The immutable nature of blockchain makes it difficult for the adversaries to alter those data. The distributed storage features can enhance accessibility and ensured security against DoS attacks.

\paragraph{ITS and Payment Services} With a massive number of vehicles on the road, it is very challenging to implement ITS and other payment services like E-tolls, intelligent parking, usages based insurance, over-the-air update, etc. Blockchain with the smart contract can securely and efficiently implement these services.

\subsection{Joint AI and Blockchain for IoV}
In this article, we propose to leverage the salient and best features of both blockchain and AI for security,  privacy  and  trust-related risks to make IoV resilient and intelligent for better performance. We also  present open  problems, challenges and requirements and potential solutions using ML and blockchain to address aforementioned issues in IoV.

\section{The need for Secure and Resilient Vehicular PKI}

Over the years, IoV has witnessed security attacks and hacks (DefCon, GeekPwn) \cite{singh2019tutorial}. The attack surface is increasing day by day because of feature integrations in the form of intelligence (sensors, ECUs, CAN), applications, services, and HetNet connectivity (V2X). Researchers have also raised concerns over open V2V communication, which can be exploited by adversaries to breach privacy by linking pseudonyms and tracking the movement \cite{9144454}. This section highlights security privacy and trust issues and key requirements to these ends.

\begin{figure}[!h]
	\centering
	\includegraphics[width=0.8\textwidth]{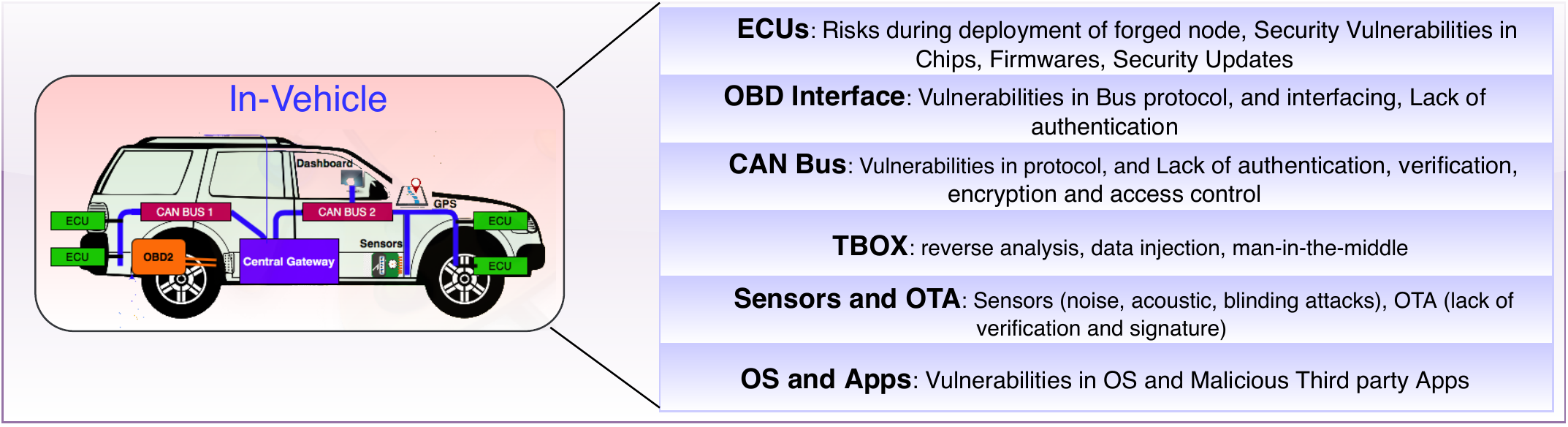}
	\caption{Security Risks at In-Vehicle System}
	\label{fig:4}
\end{figure}

\subsection{Security and Privacy Risks} 

Security risks of IoV are on the following aspects In-vehicle (Connected and Intelligent devices and applications), V2X communication, Service platforms (Edge, Fog, and Cloud), and Data. Fig \ref{fig:4} illustrates how the in-vehicle system of CAV itself is vulnerable to various types of attacks.

V2X communications have several security risks associated with them. Mainly due to flaws in the communication protocol stacks, lack of proper security management and detection mechanisms, and vulnerabilities in RATs interface and mode of communications. Some of the well-known attacks are DoS, DDoS, man-in-the-middle (MiM), Sybil, Greyhole, Blackhole, and Wormhole.

The centralized service platforms on clouds face several security challenges due to the lack of strong access control policies, certificate management, authentication mechanism, audit, intrusion detection mechanisms, etc. They are also vulnerable to DoS and DDoS attacks. 

The security of data generated and exchanged by various entities of IoV is paramount. For example, data related to location, speed, heading, brake, acceleration, tire pressure, fuel consumption, vehicle profile, driving behavior, etc. These data enable safety applications, traffic management, and other services. If these data are falsified or tampered with, it will be a severe threat to drivers and occupants' safety. Traffic management, services, and associated business applications will also be affected.

\begin{figure}[!h]
	\centering
	\includegraphics[width=0.70\textwidth]{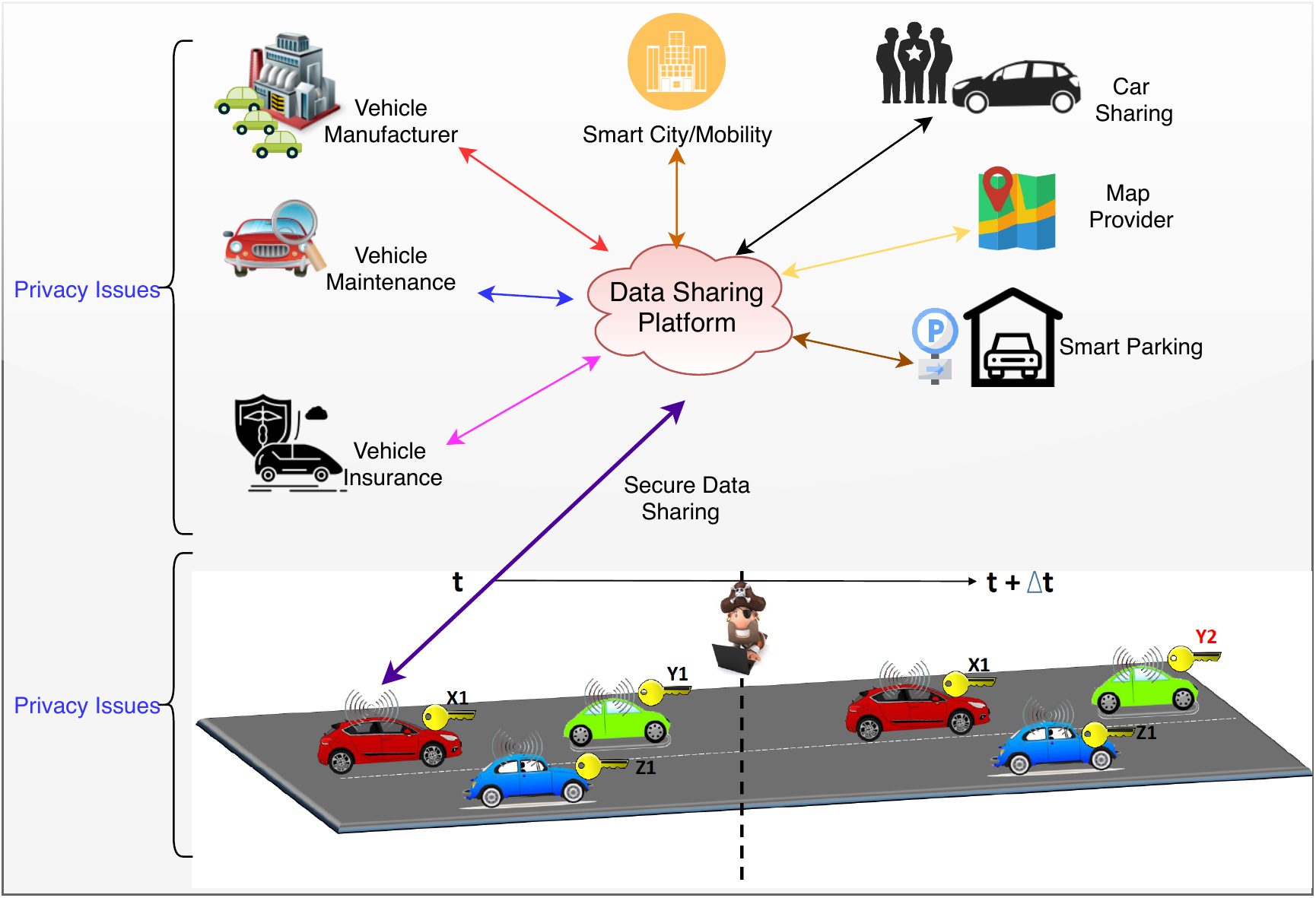}
	\caption{Privacy Risks in IoV}
	\label{fig:5}
\end{figure}

As shown in Fig. \ref{fig:5}, location tracking by dumping V2V safety messages (BSMs and CAMs) and performing syntactic and semantic linkage attacks are among the main location privacy risks. Since these safety messages are delay-sensitive and disseminated in V2V mode as plaintext (encryption not recommended due to delay), an adversary can easily eavesdrop and extract useful sensitive Spatio-temporal information. The adversary can use powerful tracking algorithms to reveal the location of the driver and occupants. Data shared with service providers and uploaded to cloud platforms via secure V2I also poses a privacy risk. For example, vehicle data are shared with manufacturers, insurance firms, service centers, map providers, and others for better services but can also be exploited to breach privacy.

\subsection{Misbehaviour and Malicious Activities}

Vehicles, roadside sensors (traffic detecting units), RSUs, network elements, and other IoV entities register with trusted authorities and receive certificates (long-term and short-term) and cryptographic materials (key pairs). These materials and certificates are used in communication to ensure confidentiality (V2I/I2V, I2I), integrity, authenticity through the signing and verification process. On the other hand, these certifications and materials do not safeguard the IoV from internal threats posed by compromised, malevolent, or misbehaving entities. For example, a vehicle in the network can broadcast misleading and false information about its kinematics, creating chaos and jeopardizing road safety. Similarly, a compromised or faulty traffic detection unit (TDU) can push wrong traffic info (congestion at that road even when no traffic) to the traffic authority (TA). When this incorrect info is propagated to commuters by TA, it may disrupt the entire traffic in that zone. Such types of actions are known as misbehavior in IoV. Malicious entities can transmit misleading information on purpose, while malfunctioning or compromised entities can unintentionally convey wrong info. These attackers are considered insiders and active attackers since they have the necessary certificates and cryptographic materials to communicate in an IoV. By getting access to their vehicle's CAN bus  (since the protocol is vulnerable), these attackers can change the payload of outgoing safety and awareness messages. It's also conceivable for attackers to use a MiM and replay attack to change sensor data. \cite{kamel2020simulation}.

\begin{figure}[!h]
	\centering
	\includegraphics[width=0.42\textwidth]{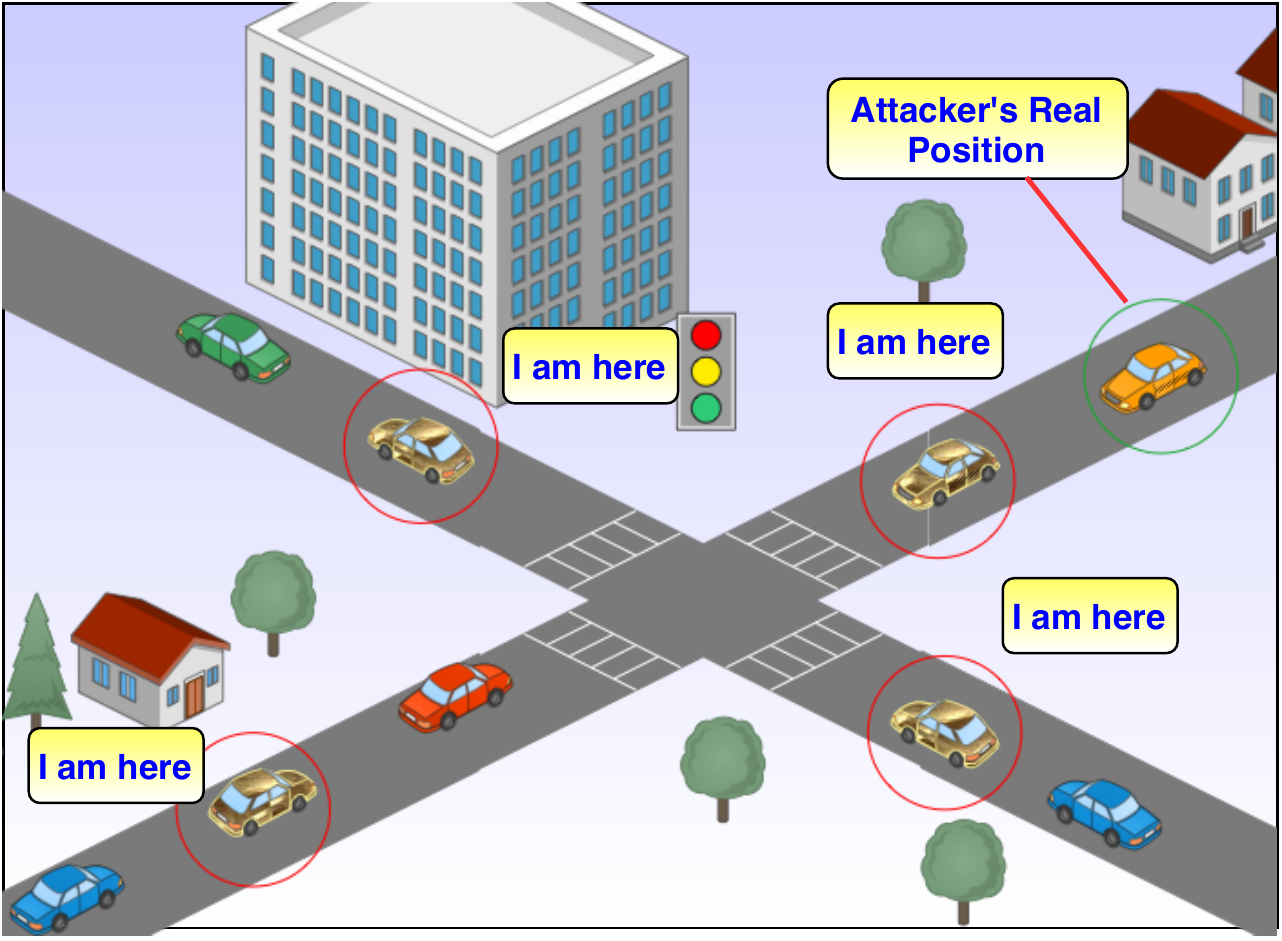}
	\caption{Misbehaviour (Position Falsification Attack) in IoV.}
	\label{fig:6}
	\vspace{-1em}
\end{figure}

Fig. \ref{fig:6} depicts an example of one such type of misbehavior known as position falsification attack. The vehicle misbehaves by changing the payload of the safety message to produce bogus positions and broadcasting them through V2V.  As depicted in Fig. \ref{fig:6}, When the misbehaving node is close to the position of the intersection, it is overwhelming false location information and creating confusion for other vehicles \cite{kerrache2016trust}. Such types of misbehavior can have disastrous consequences, which may pose a risk to other nodes nearby. Because the message is generated by an insider (authorized peer), it gets verified and accepted for processing by other receiving vehicles that are part of the same authentication system. As a result, a wide range of safety-related applications that rely on actual position data may be adversely affected.

\subsection{Security, Privacy and Trust Requirements}
Avoiding those IoV security and privacy risks and misbehavior may lead to severe consequences. Thus, guaranteeing security and privacy and dealing with misbehavior (trust management) are vital requirements of the IoV for its acceptability. Some of those key requirements are summarized below.

\subsubsection{Security Requirements}
There is a need for strong security solutions and mechanisms at different levels of IoV. 

\textbf{In-vehicle Security:} The in-vehicle system demands advanced hardware security solutions and modules to secure OBD2, ECUs, T‐BOX, event recorder, storage, onboard units, sensors (LiDAR, RADAR, GPS, Cameras, TMPS, Gyroscope), etc. There is a need for secure design and development of software (OS and Apps) and firmware of the vehicles. Over-the-Air update of software and firmware needs to be secure and immutable. The in-vehicle communication modules such as CAN, FlexRay, LIN, Ethernet, Bluetooth, etc., need to be secured. The core security functions such as encryption, access control, authentication, integrity check, signing, and verification need to be utilized. Besides, we need to have a strong firewall and anomaly detection mechanisms for the in-vehicle system. 

\textbf{V2X Security:} Securing the HetNet V2X communications (DSRC, cellular, VLC, Wi-Fi, etc.)  will be challenging. However, all forms of V2X communication need to fulfill the following security requirements. (1) Advanced cryptographic approaches must be used to fulfill security requirements such as availability, access control, authentication, confidentiality, non-repudiation, integrity, and data verification at different planes of IoV. These should be used in such a way that data from an authenticated entity can be transmitted to other entities(s) of the system securely without any alteration or tampering. (2) The consistency and rationality of the communication data need to be protected. (3) The security of data stored at the cloud, fog, and edge and their secure exchange (offloading, uploading) with infrastructure and vehicles must be guaranteed. 

\textbf{Service Platforms:} Deployed security capabilities for cloud platforms need to be strengthened for IoV services. Essential security requirements are as follows. (1) The data access control mechanisms need to be enhanced and highly secured. (2) Need to ensure secure cross-cloud (OEMs, government agencies, service providers) or fog interactions. (3) There is a need to establish trust among service providers to determine the level of data sharing and exchange. (4) Strong authentication mechanisms, security policies, and firewalls must be in place to secure user and vehicle data stored in the cloud from malicious users and hackers. 

\subsubsection{Privacy Requirements:}
A strong privacy protection mechanism is required to prevent attackers from exploiting private data about the vehicle, the driver, and the occupants. In IoV, there are three distinct kinds of privacy protection mechanisms: (1) Vehicle's actual identity should not be revealed. (2) Vehicle's location should not be tracked, and (3) All data exchanged in IoV are protected against privacy. The following are the IoV's most important privacy requirements: \cite{boualouache2017survey}:

\textbf{Minimum disclosure:} The information about the user that is absolutely necessary for IoV functionalities should be revealed, and it should be kept minimal.

\textbf{Anonymity:} The usage of pseudonyms provides anonymity, which is one of the most common ways to preserve privacy. For example, the messages delivered by a vehicle should be anonymous within a set of subjects, such as potential vehicles, and should not reveal the vehicle's real identity.

\textbf{Unlinkability:} The use of a pseudonym aids authentication while keeping the true identity hidden. However, if a pseudonym is used in the same context for a long time, it will become linkable. Therefore, to achieve the unlinkability of pseudonyms, a set of pseudonyms is used. To avoid linkability in the given context, these pseudonyms must be changed over time.

\subsubsection{Trust Management}In the IoV, trust management is a crucial concept for identifying and revoking malicious and misbehaving nodes. The adoption of proper trust models helps in determining the trustworthiness of the message and its sender. The following are the major requirements for trust management in IoV:
\textbf{Misbehavior Detection:} One of the most important aspects of trust management is detecting malicious and misbehaving nodes. Its goal is to keep an eye on the system in order to spot any potentially misbehaving nodes and keep the IoV from straying from its normal course. The technique for detecting misbehavior works in four stages \cite{kamel2020simulation}: Local detection of the misbehaving entity, reporting of the misbehavior to the central authority or Misbehavior Authority (MA), assessment by the MA to establish whether the entity is truly misbehaving or simply faulty, and lastly reaction by the registration authority (RA) to protect the system.

\textbf{Accountability and Traceability:} To establish accountability, the specified action should be unambiguously assigned to an individual entity via a fair methodology or protocol. Traceability can help with this in the IoV. Only authorized and highly trusted authorities should be able to trace a pseudonym and link it to the user's true identity. Under some specific instances, such authorities must attempt to trace or map a pseudonym to a real identity. Only MA, for example, should be permitted to do tracing in order to identify the real misbehaving entity.

\textbf{Revocation and CRL Distribution:} When misbehavior is detected, trust management in the IoV should have a fair revocation process that can act effectively. Some reputation-building processes with a reward and penalty system should also be implemented. If the misbehavior is deliberate, the corresponding vehicle should be penalized, and its reputation score should be deducted. Such misbehaving entities should be issued a warning. After crossing the specified threshold, an active revocation (revocation of current certificates) or passive revocation (not able to request more certificates) should be invoked. 

\textbf{Scalability:} Since IoV is very dynamic (varying density), it can experience high traffic during peak hours. Thus, the system deployed for trust management must be scalable to deal with a large number of nodes.

\textbf{Decentralization:} Centralized trust management systems for IoV face several challenges in terms of performance, scalability, robustness, fault-tolerance, privacy, and security \cite{9142395}. Thus, there is a need for fully decentralized trust management for highly dynamic and distributed IoV.

\subsection{Standardization Efforts}

The security standard IEEE 1609.2 specifies security, privacy protection, and trust management in the connected vehicle protocol stack of the USA. The security standards ETSI TS 102 940 V 1.2.1 and ETSI TS 102 941 specify the security, privacy, and trust in the Cooperative-ITS protocol stack of Europe. Various other efforts are underway by SAE (J3061), ITU (FG and SG17), ISO/TC22, CEN, and 3GPP (C-V2X). The IEEE and ETSI standards recommend using pseudonyms for privacy. However, determining when and how to change pseudonyms is still a work in progress. Misbehavior detection and trust management are still in the initial phases of development under those standards.

\section{AI and Blockchain Based Framework}
The integration of AI and Blockchain in IoV can help to a great extent for its security privacy and trust management. In this section, the AI and Blockchain driven solution for the same is discussed through a generic illustration (Fig. \ref{fig:7}).

\subsection{ML for Intrusion and Misbehavior Detection}

Machine Learning (ML) paradigm facilitates building models that can anticipate the expected behavior by learning from experience. The data-driven approaches enable the machine to learn the relationship, which are hidden, and models for misbehavior detection can leverage this to find deviations. The deep learning models are found suitable in approximating such relationships.

\begin{figure}[!h]
	\centering
	\includegraphics[width=.9\textwidth]{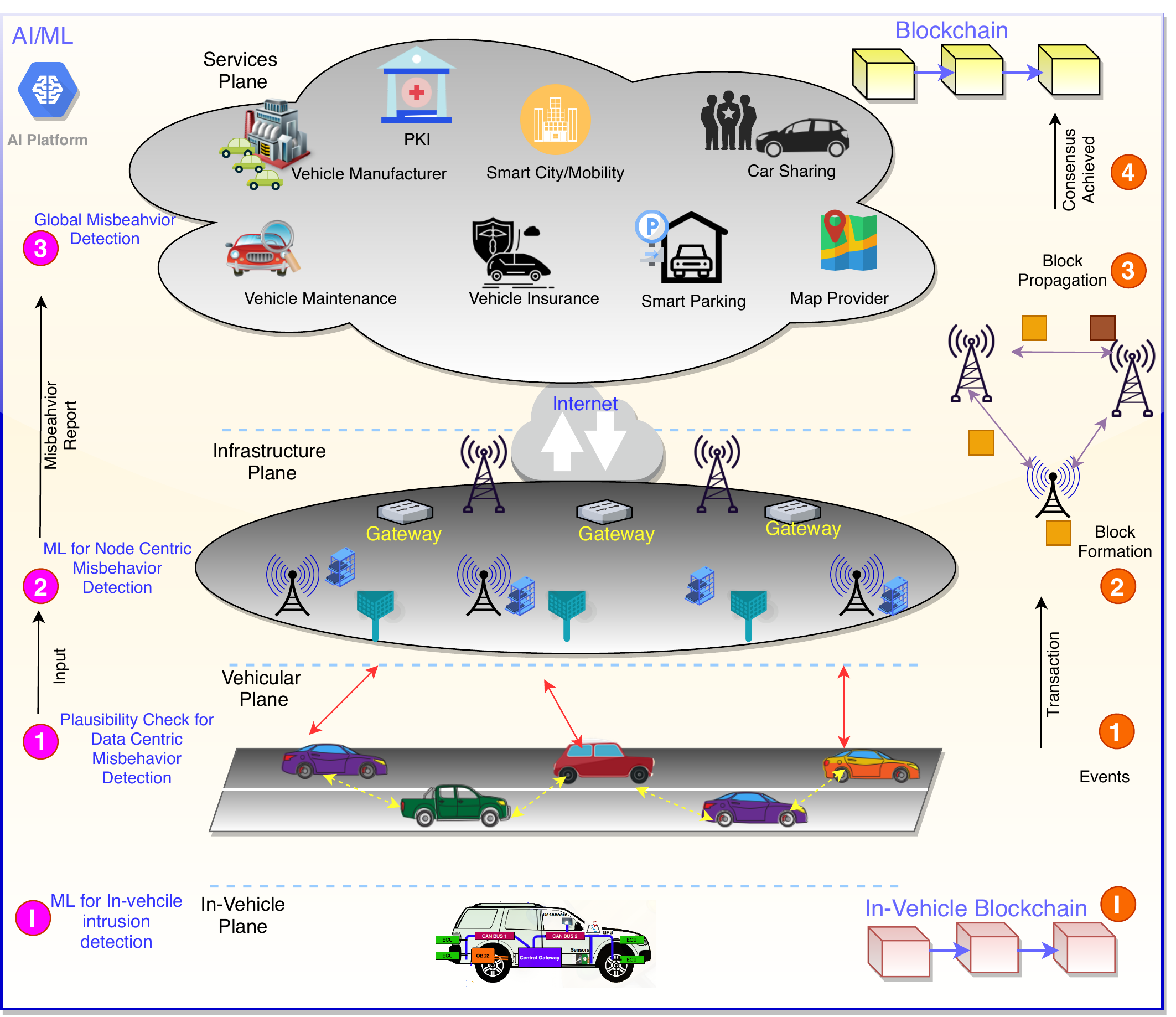}
	\caption{AI and Blockchain for Security Privacy and Trust in IoV (a Generic View)}
	\label{fig:7}
\end{figure}

Over the years, various ML-based solutions have been proposed for intrusion, anomaly, and misbehavior detection in VANETs. As shown in Fig. \ref{fig:7}, the ML can help to detect intrusions in in-vehicle network and misbehavior in vehicular as well as infrastructure plane of IoV. 

For an in-vehicle network, the intrusion is detected by comparing the current action against the expected actions. The deep learning models can extract the features from the behavior patterns of the nodes and look for the existing feature map to predict the malicious activities of node(s). The training of deep learning models through probability-based feature vectors will also be helpful in detecting intrusions. 

For misbehavior detection, the power of ML can be utilized to cope up with insider attacks. First, data-centric mechanisms (basic plausibility checks) can be used to check message consistency in the vehicular plane. For example, to check if two consecutive safety messages coming from the same node have plausible separating distance or not. Based on the results, the node-centric approach powered by deep learning models at the infrastructure plane can be invoked to detect the misbehaving or faulty entity.  

\subsection{Blockchain for Security, Privacy and Trust}

In IoV, a blockchain can be deployed at different planes such as in-vehicle, the RSU or edge plane, and at service plane (standalone or hierarchical way). The planes that can support resources required for the transaction processing and procedure for forming new blocks can be considered. The private blockchain is found to be more suitable for such contexts. 

\textbf{Security:} To secure complex cyber-physical system against hacks and attacks (mentioned above) of in-vehicle, there can be a separate private blockchain. The powerful nodes, such as gateways and switches, can be considered as blockchain nodes. The consensus protocols can keep nodes synchronized with each other. All communications across ECUs can be considered as transactions. The smart contracts can be used, which can invoke ML for anomaly or intrusion detection, and if detected, those transactions can be dropped (depending on the defined danger level) and should not be considered in blocks. Similarly, to secure vehicular plane, blockchain deployed at the edge can be helpful. Participating vehicles generate a series of transactions for various events happening on the road, such as applying the harsh brake, detecting slippery roads, etc. Transactions generated by vehicles on the road are sent to the associated RSU. The suspicious transaction can be dropped at the edge level (smart contract can be used with local detection events). An RSU stores the transaction in its memory pool and simultaneously propagates the transaction among the peer RSUs. The block formation task is entrusted to the RSUs (edge nodes). RSUs pick a random set of transactions from their pool of received transactions and perform a mathematical operation on the selected transactions in order to form a block. Once a block is formed, an RSU propagates the formed block to its network of RSUs. Each RSU in the network verify the received block and accept it if it is a valid one. 

\textbf{Privacy and Trust Management:} The Public Key Infrastructure (PKI)-based pseudonym authentication mechanism can be used in the deployed blockchain at the cloud with a certain decentralization for better certificate and pseudonym management and their efficient distribution. The location privacy can be preserved with masquerading concepts over the existing pseudonym change mechanism, which need to be backed up by proof-of-claim and casual dependency to ensure traceability (by trusted authority) and non-repudiation \cite{9144454}. To preserve the privacy in data sharing, smart contracts can be used to sign agreements between the vehicle (owner) and service providers. The blockchain can be invoked to ensure that signed smart contacts do not get tampered with. It also enforces proper access control to implement secure and authentic data accessibility and services. The reputation management can also be implemented using smart contracts to penalize the misbehaving vehicles and revoke the certificate or rewarding them (better trust value) based on the report received from the infrastructure plane. The power of AI can be utilized to ensure accountability in the system.

\section{Open Challenges and Research Perspectives}

This section of the article discusses challenges and research opportunities in AI and Blockchain enabled IoV.

\subsection{Performance: Latency, Throughput, Energy and Scalability}

Blockchain and AI adoption to IoV may require handling massive data and transaction processing in a highly dynamic environment. Thus maintaining performance to the allowable latency threshold, throughput, and energy are major concerns \cite{mollah2020blockchain}. The acceptable latency for safety and non-safety applications is in the range of 100ms to 1 second. For delay-sensitive safety applications, maintaining latency in milliseconds for the entire process of communication, transaction processing, consensus, and block formation will be the biggest challenge. Similarly, maintaining throughput, minimizing energy consumption (mainly electric vehicles), and dealing with a large number of nodes in peak hours of IoV are the biggest challenges of AI and Blockchain integration.

\subsection{Security, Privacy, Transparency} AI and Blockchain integration into IoV may result in a tradeoff between Security and Privacy and Privacy and Transparency. Dealing with such a scenario will be very challenging. Security flaws have also been reported to Blockchain technology structure, peer-to-peer system, and application in the past. The poisoning attack in AI/ML algorithms may mislead the decision making. Ensuring anonymity of users’ identity and providing unlinkability of transactions are two major concerns from a privacy perspective. AI powers can be misused to link the pseudonym identity and transactions. Thus security, privacy, and transparency need to be widely explored prior to AI and Blockchain integration. 

\subsection{Heterogeneity, Deployment Strategies and Standardization} IoV is heterogeneous at access networks, core network technologies, data formats, etc.  Thus, dealing with HetNet and different data formats and semantics are very challenging. AI and Blockchain deployment strategy used in one scenario of IoV may not be suitable for other scenarios. For example, Proof-of-Work (PoW) may be ideal for delay-tolerant non-safety applications where security is a major concern but not suitable for delay-sensitive safety applications. Similarly, if the complex AI models (deep learning) used with PoW may lead to high computation, overhead, and delay. Thus, the selection of a proper combination of AI and Blockchain is challenging. The interoperability and standardization of the AI and Blockchain deployment strategy for IoV remain the most significant concerns to be addressed. 

There are enormous research opportunities to tackle the challenges mentioned above. How AI and Blockchain can complement each other in IoV is not yet explored much. More efforts need to be put in to fulfill the QoS, security, privacy, and trust requirement of IoV. Making the Blockchain solution for IoV quantum secure using advanced cryptography such as lattice-based and multivariate mechanisms is one of the hot topics. How private blockchain can be used to secure the complex in-vehicle CPS is another area of research. Designing and developing the adversarial machine learning models for IoV is another open research area. The HetNet scenario of IoV itself opens enormous research opportunities to develop a common, interoperable, and standard security framework. 


\section{Conclusion}
For public acceptance and successful implementation of IoV, security, privacy protection, and trust management are critical aspects. The Blockchain and AI-based solutions can be integrated with the widely used security infrastructure, namely PKI for IoV, such as Security certificate management System (SCMS) in the USA and ETSI TS 102 940 and 941 in Europe to achieve the desired level of protection against security and privacy threats. The trust management process can also be enhanced by ML-based misbehavior detection and inheriting the blockchain-based incentivization approach for revocation (reward and penalty). The AI can help detect anomalies and intrusions at different planes of IoV when invoked through the smart contract logic. Blockchain as a decentralized solution can prevent malicious activities and resolve fault-tolerance and scalability issues. This article discussed the role of AI and Blockchain in IoV separately. Then highlighted the security, privacy, and trust issues in the IoV. How AI and Blockchain can help in dealing with security, privacy and trust issues was discussed through a framework. Furthermore, this article has discussed several open challenges and possible future research directions to tackle them.

\ifCLASSOPTIONcaptionsoff
  \newpage
\fi

\bibliographystyle{IEEEtran}
\bibliography{ref}

\begin{IEEEbiographynophoto}{Pranav Kumar Singh}
is working as an Assistant Professor in the Department of Computer Science and Engineering, Central Institute of Technology Kokrajhar, India. Contact him at snghpranav@gmail.com
\end{IEEEbiographynophoto}

\begin{IEEEbiographynophoto}{Sukumar Nandi} is a Senior Professor with the Department of Computer Science \& Engineering, Indian Institute of Technology Guwahati, India. Contact him at sukumar@iitg.ac.in 
\end{IEEEbiographynophoto}

\begin{IEEEbiographynophoto}{Sunit Kumar Nandi}
	is currently pursuing his Ph.D. in the Department of Computer Science and Engineering, Indian Institute of Technology Guwahati, India. Contact him at sunitnandi834@gmail.com
\end{IEEEbiographynophoto}

\begin{IEEEbiographynophoto}{Uttam Ghosh}
	is working as an Assistant Professor of Practice in EECS, Vanderbilt University, Nashville, Tennessee, USA. Contact him at ghosh.uttam@ieee.org 
\end{IEEEbiographynophoto}

\begin{IEEEbiographynophoto}{Danda B. Rawat}
	is a Professor at the Department of Electrical Engineering and Computer Science, Howard University, Washington, DC, USA. Contact him at db.rawat@ieee.org
\end{IEEEbiographynophoto}

\end{document}